\documentclass[prl,twocolumn]{revtex4}
\usepackage{epsfig}
\usepackage{amsmath,amssymb}

\begin{document}

\title{The Limits of Quintessence}
\author{R.~R. Caldwell} 
\affiliation{Department of Physics \& Astronomy, Dartmouth College, 
Hanover, NH 03755}
\author{Eric V.\ Linder} 
\affiliation{Physics Division, Lawrence Berkeley Laboratory,
Berkeley, CA 94720}

\date{\today}

\begin{abstract}

We present evidence that the simplest particle-physics scalar-field models of dynamical
dark energy can be separated into distinct behaviors based on the acceleration or
deceleration of the field as it evolves down its potential towards a zero minimum. We
show that these models occupy narrow regions in the phase-plane of  $w$ and $w'$, the
dark energy equation-of-state and its time-derivative in units of the Hubble time. 
Restricting an energy scale of the dark energy microphysics limits how closely a scalar
field can resemble a cosmological constant. These results, indicating a desired
measurement resolution of order  $\sigma(w')\approx (1+w)$, define firm  targets for
observational tests of the physics of dark energy. 

\end{abstract}

\maketitle

Observations and experiments at the close of the 20th century have transformed our
understanding of the physics of the Universe. A consistent picture has emerged
indicating that nearly three-quarters of the cosmos is made of ``dark energy'' --- some
sort of gravitationally repulsive material  responsible for the accelerated expansion
of the Universe (for reviews see  \cite{rmp,padma,Caldwell:2004ze}). Proposals for the
dark energy include Einstein's cosmological constant ($\Lambda$), or a dynamical field
such as quintessence. Here we show how scalar field dynamics separates into distinct
behaviors which, through future cosmological measurements, can reveal the nature of the
new physics accelerating our universe.

Einstein's cosmological constant ($\Lambda$) is attributed to the  quantum zero-point
energy of the particle physics vacuum, with a  constant energy density $\rho$, pressure
$p$ and an equation-of-state  $w\equiv p/\rho = -1$. In contrast, quintessence is a
proposed time-varying, inhomogeneous field with a spatially-averaged equation-of-state
$w > -1$ \cite{Ratra:1987,Wetterich:1988,Frieman:1995pm,Coble:1996te,Caldwell:1997}.
The simplest physical model consists of a scalar field, slowly rolling in a potential
characterized by an extremely low mass. (This is similar to inflation, the period of
accelerated expansion in the early universe, but at an energy scale many orders of
magnitude lower.) Since a scalar field evolving in a very shallow potential may be
indistinguishable from a $\Lambda$, the task of elucidating the physics of dark energy
becomes difficult if observations continue to find that $w$ is close to $-1$, {\it
e.g.} \cite{knop,riess,seljak}. In this letter, we examine the likely behavior of
scalar fields and characterize them into two distinct classes, based on their evolution
in the $w-w'$ phase space. These results should help define targets for observational
and experimental tests of the physics of dark energy.

Our approach is a new take on a familiar system, the scalar field.  By  emphasizing the
dynamics, we discover restricted regions of the trajectories  of canonical scalar field
models in ``position'' and ``velocity'' ---  the value of the equation-of-state ratio
$w$ and its time variation $w'$.  While there is a myriad of scalar field models
motivated by particle physics beyond the standard model, this treatment allows a broad,
model-independent  assessment of a quintessence scalar field slowly relaxing in a
potential. 

\begin{figure}[b]
\includegraphics[scale=1.0]{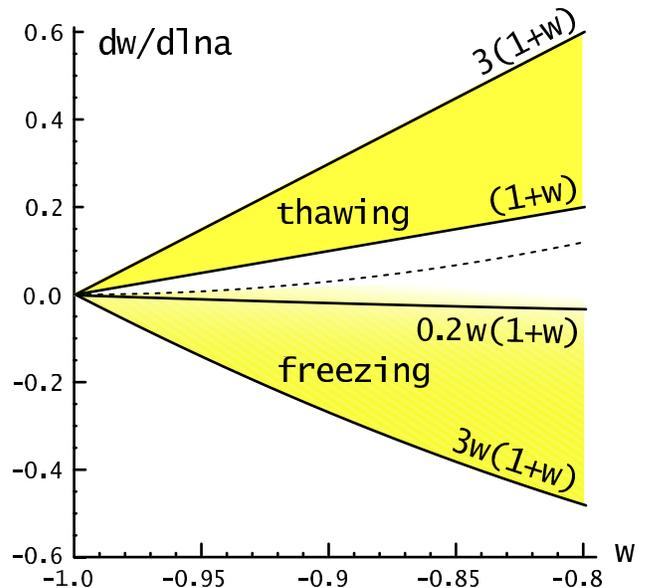}
\caption{The $w-w'$ phase space occupied by thawing and freezing fields is indicated by
the shaded regions. No strong constraints on this range of dark energy properties exist
at present. The fading at the top of the freezing region indicates the approximate
nature of this boundary. Freezing models start above this line, but pass below it by a
red shift $z \sim 1$. The short-dashed line shows the boundary between field evolution
accelerating and decelerating down the potential. Future cosmological  observations
will aim to discriminate between these two fundamental scenarios.} 
\label{fig.shaded}
\end{figure}

The physics is straightforward: the field $\phi$ will seek to roll towards the minimum
of its potential $V$, according to the Klein-Gordon equation $\ddot\phi+3 H\dot\phi=
-dV/d\phi$. The rate of evolution is driven by the slope of the potential and damped by
the cosmic expansion through the Hubble parameter $H$. The average energy density and
pressure are $\rho=\dot\phi^2/2+V,\,p=\dot\phi^2/2-V$ so that a field stuck in a local,
non-zero minimum of the potential has $w=-1$. To distinguish from an effective
cosmological constant, however, we will only consider cases in which the field is
evolving towards a zero minimum.

In perhaps the simplest such scenario, the field has been frozen by Hubble damping at a
value displaced from its minimum until recently, when it starts to roll down to the
minimum. We call these ``thawing'' models.  At early times the equation-of-state ratio
is $w\approx -1$, but grows less negative with time as $w'\equiv\dot w/H>0$. Since the
Hubble damping limits the scalar field acceleration, $\ddot\phi < \dot\phi/t \approx
(3/2)H\dot\phi$, then the equation of motion implies such models will lie at
$w'<3(1+w)$ in the phase plane. The scalar field dynamics suggest a lower bound, too,
due to the fact that dark energy is not entirely dominant today, with a fractional
energy density $\Omega_{de} \lesssim 0.8$. A study of several classes of thawing
models, such as a pseudo Nambu-Goldstone boson (PNGB) \cite{Frieman:1995pm} or
polynomial potentials, indicates the bound $w'>(1+w)$. These simple bounds are valid
for $(1+w)\ll 1$, and so $w \lesssim -0.8$ is a practical limit of applicability.

We have analyzed the following potentials for thawing behavior: Concave potentials with
$V=M^{4-n}\phi^n$ are ubiquitous, and we have allowed for continuous values of the
exponent $n > 0$. The motivation for cases $n<2$ is not as straightforward, although
$n=1$ has been considered \cite{Kallosh:2003}. Exponential potentials are typical for
moduli or dilaton fields, {\it e.g.\/} \cite{Barreiro:1999zs}, with $V=M^4 \exp(-\beta
\phi/M_{P})$ where $M_P \approx 10^{19}$~GeV is the Planck energy. To avoid scaling,
which would not provide for the cosmic acceleration, we restrict $\beta < \sqrt{24
\pi}$ \cite{Ferreira:1997hj}. PNGBs, like a dark energy axion, have $V = M^4
\cos^2(\phi/2 f)$ where $f$ is a symmetry restoration energy scale.  We have not
included the case $f \ll M_P$ since the field rapidly evolves to $w\to 0$ unless the
initial conditions are finely tuned to keep the field balanced upon the top of the
potential maxima and maintain $1+w \ll 1$.

A second scenario consists of a field which was already rolling towards its potential
minimum, prior to the onset of acceleration, but which slows down  and creeps to a halt
as it comes to dominate the universe.  For these  ``freezing'' models, initially $w >
-1$ and $w' < 0$. These are essentially tracking models \cite{Zlatev:1998tr}, but may
be described more generally as vacuumless fields (in the sense that the minimum is
attained as $\phi\to\infty$) or runaway potentials characterized by a potential with
curvature that slows the field evolution as it rolls down towards the minimum. It
follows \cite{Steinhardt:2000} that there is some value of the field beyond which the
evolution is critically damped by the cosmic expansion, whence the field is frozen
(but, like a glacier \cite{Steinhardt:2000}, continues to move) and $w \to -1,\, w'\to
0$. The deceleration of the field is limited by the steepness of the potential, roughly
$\ddot\phi>dV/d\phi$, leading to the lower bound $w'>3 w(1+w)$.  Investigation of a
variety of scalar field models leads to a less definite upper bound $w'\lesssim 0.2
w(1+w)$ since a red shift $z\sim 1$ but evolving beyond $w' \lesssim w(1+w)$ by the
present.  Again, $w \lesssim -0.8$ is a practical limit of applicability of these
bounds.

We have analyzed the following tracker potentials for their freezing behavior: $V =
M^{4+n}\phi^{-n}$ and $V= M^{4+n}\phi^{-n} \exp(\alpha \phi^2/M_P^2)$ for $n>0$
\cite{Binetruy:1998rz,Brax:1999gp,Masiero:1999sq,Copeland:2000vh,delaMacorra:2001ay}.
The latter has an effective cosmological constant, but has been widely studied and so
we consider it nonetheless, provided the field is closer to the origin than the
non-zero minimum.  Proposed tracker models such as $V=M^4 \exp(M_P/\phi)$  do not have
a zero minimum, and $V=M^4 (\exp(M_P/\phi)-1)$ does not achieve $w \lesssim
-0.8,\,\Omega_{de} \lesssim 0.8$. Other functions have been proposed as tracking
potentials, but lack a firm basis in particle theory.

\begin{figure}[b]
\includegraphics[scale=1.0]{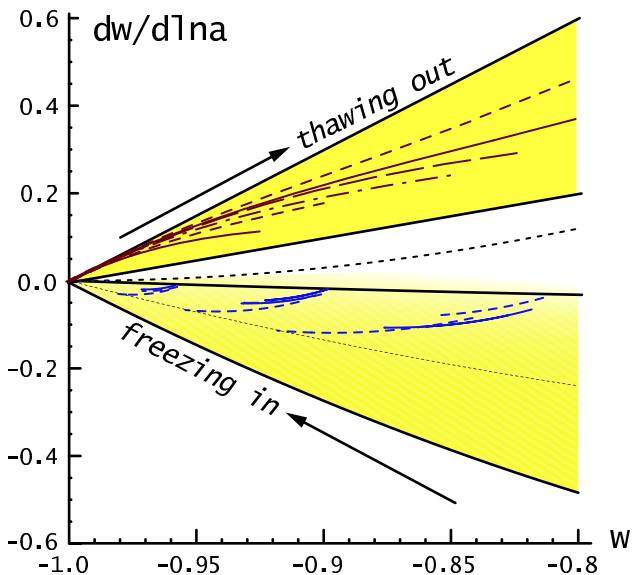}
\caption{The evolutionary tracks in $w-w'$ phase space are shown for a  variety of
particle physics models of scalar fields. The two broad classes are clear: those that
initially are frozen and look like a cosmological constant, starting at $w=-1$, $w'=0$,
and then thaw and roll to $w'>0$, and those that initially roll and then slow to a
creep as they come to dominate the Universe. The sample of thawing models shown have
potentials $V\propto \phi^n$ for $n=1,\,2,\,4$ (short-, dot-, and long-dashed curves)
and a PNGB with $V\propto \cos^2(\phi/2f)$ (solid curves). The right-most point of the
tracks corresponds to the present. For variety, the $n=4$ model has $\Omega_{de}=0.6$,
and the $n=1$ model ending at $w=-0.8$ has $\Omega_{de}=0.65$. All other models end
with a fractional energy density $\Omega_{de}=0.7$. The sample of freezing models shown
have potentials $V\propto \phi^{-n}, \, \phi^{-n}{\rm e}^{\alpha\phi^2}$ (solid and
dashed curves). The line $w'=1.5w(1+w)$ indicated by the light, dotted line is a
possible lower bound on the freezing models. The left-most point of the tracks
corresponds to the present; the right-most point is at $z=1$.  For variety, upper and
lower close pairs of curves have $\Omega_{de}=0.7,\,0.8$ respectively. All other models
end with a fractional energy density $\Omega_{de}=0.7$. } 
\label{fig.models}
\end{figure}

These distinct, physically motivated ``thawing'' and ``freezing'' behaviors  are
illustrated in Figure~\ref{fig.shaded}, while several examples of specific models are
presented in Figure~\ref{fig.models}. It would be quite useful to determine if one of
these classes of physics phenomena is responsible for the dark energy accelerating our
universe. We see that to distinguish thawing from freezing, a measurement resolution of
order $\sigma(w') \approx (1+w)$ is required.  

The question of the absolute level of deviation of $w$ from $-1$, i.e.\  the
distinction from Einstein's cosmological constant, is less tractable. Certainly, one
can obtain a scalar field solution, however unrealistic, at any given point in the
$w-w'$ phase space. Even for the thawing and freezing models, parameters may be finely
tuned to keep $1+w$ arbitrarily close to zero within the shaded regions of
Figure~\ref{fig.shaded}.  

If the scalar field is prohibited from attaining values exceeding the Planck scale,
lest quantum gravitational effects dominate, then there is a lower bound on $1+w$.
Defining a characteristic scale $E \equiv |V/(dV/d\phi)|$ we demand $E<M_P$. Next, for
a field rolling down its potential, we can express the scalar field equation of motion
as $$w'=-3(1-w^2) + (1-w)\frac{M_P}{E}\sqrt{\frac{3}{8 \pi}\Omega_{de}(1+w)}.$$ 
Taking $\Omega_{de}\approx 0.7$ then thawing models must satisfy $1+w \gtrsim 0.004$
whereas for freezing models $1+w \gtrsim 0.01$; this may be the limit of quintessence.
These margins correspond to a $\sim 0.2\%$ difference from a $\Lambda$ cosmology in
distance to redshift $z=1$. Such an absolute precision goal is clearly extraordinarily
challenging.  

Note that early universe inflation can similarly approach pure exponential expansion,
with its deviations broadly characterized by dynamics into models tilted to prefer
large-scale or small-scale power, and important implications in the distinction
\cite{kinney}. The structure we find in the canonical scalar-field phase-plane based on
simple physical considerations may prove useful, and we have here presented firm
targets for a basic test of dark energy. The language is different from inflation for
two reasons: the dark energy need not be rolling as slowly as the inflaton, and the
dark energy is not totally dominant, unlike the inflaton.

Charting the late-time cosmic evolution -- through Type Ia supernovae distances, weak
gravitational lensing probes of large-scale structure evolution, distance ratios from
baryon acoustic oscillations in galaxy clustering, {\it etc} -- is the subject of
intense investigations. As a gauge of the requisite resolution, a 1\% variation in
luminosity distance to redshift $z=1$ distinguishes between: $\Lambda$ and
$w=-0.95,\,w'=0$; models which evolve along the top and bottom of the thawing region
out to $w=-0.8$; models which evolve along the top and bottom of the freezing region in
to $w=-0.95$. The goal of making the fundamental physics distinction between the
thawing and freezing regions is challenging but achievable in the next generation of
experiments \cite{omnibus,Jarvis:2005ck,Linder:2004qc,Upadhye:2004hh} if the dynamics
is sufficiently apparent. In the case $1+w\gtrsim 0.05$, dedicated dark energy
experiments now being designed, such as the Joint Dark Energy Mission, will probe
cosmology with sufficient  accuracy to be able to decide the issue. 

This will probably not be the final word on dark energy ({\it cf.}\ \cite{Maor:2002rd}),
but if the answer is not consistent with a cosmological constant then the rewards are
obvious in discovering new physics beyond our current standard models.  It is
interesting to note that the fate of the Universe is very different for the case of a
thawing field, as the acceleration is temporary, as compared to a freezing field, for
which the acceleration continues unabated. If the result lies outside the two
phase-space regions categorized here then we may have to look beyond simple
explanations, perhaps to even more exotic physics such as a modification of Einstein
gravity.  
 
\acknowledgments
This work is supported by NSF AST-0349213 at Dartmouth, and by DOE 
AC03-76SF00098 at LBL.  EL thanks Dartmouth for its hospitality. 
 


\end{document}